\documentclass[reprint,bibnotes,amsmath,amssymb,aps,longbibliography]{revtex4-1}

\usepackage{graphicx}
\usepackage{dcolumn}
\usepackage{bm}
\usepackage{hyperref}
\usepackage{csquotes}
\hypersetup{colorlinks=true,linkcolor=red,citecolor=red}
\usepackage{float}
\usepackage[dvipsnames]{xcolor}
\usepackage{gensymb}

\usepackage{prettyref}%
\newcommand{\pref}[1]{\prettyref{#1}}%
\newrefformat{fig}{Fig.~\ref{#1}}%
\newrefformat{tab}{Table~\ref{#1}}%
\newrefformat{sec}{Sec.~\ref{#1}}%
\newrefformat{app}{App.~\ref{#1}}%
\newrefformat{eq}{Eq.~(\ref{#1})}%

\DeclareUnicodeCharacter{2212}{-}

\begin{document}

\title{Effect of chemical disorder on the magnetic anisotropy in L$1_0$ FeNi from first principles calculations}

\author{Mayan Si}
\author{Ankit Izardar}
\author{Claude Ederer}
\email{claude.ederer@mat.ethz.ch}
\affiliation{Materials Theory, ETH Z\"urich, Wolfgang-Pauli-Strasse 27, 8093 Z\"urich, Switzerland}

\date{\today}

\begin{abstract}
We use first principles calculations to investigate how deviations from perfect chemical order affect the magneto-crystalline anisotropy energy (MAE) in L$1_0$ FeNi. We first analyze the local chemical environment of the Fe atoms in various partially ordered configurations, using the orbital magnetic moment anisotropy (OMA) as proxy for a local contribution to the MAE. We are able to identify a specific nearest neighbor configuration and use this ``favorable environment'' to successfully design various structures with MAE higher than the perfectly ordered system. However, a systematic analysis of the correlation between local environment and OMA using smooth overlap of atomic positions (SOAP), indicates only a partial correlation, which exists only if the deviation from full chemical order is not too large, whereas in general no such correlation can be identified even using up to third nearest neighbors. Guided by the observation that the identified ``favorable environment'' implies an Fe-rich composition, we investigate the effect of randomly inserting additional Fe into the nominal Ni planes of the perfectly ordered structure. We find that the MAE increases with Fe content, at least up to 62.5\% Fe. Thus, our study shows that the perfectly ordered case is not the one with highest MAE and that an increased MAE can be obtained for slightly Fe-rich compositions.
\end{abstract}

\maketitle

\section{\label{sec:Intro}Introduction}

Due to the huge and strongly increasing demand of permanent magnets for, e.g., applications in electrical power generation and conversion, there is great interest in new magnetic materials, in particular those containing cheap and abundant elements with no or only small amounts of rare earth elements~\cite{Gutfleisch_et_al:2011,Ronning/Bader:2014}. L$1_0$-ordered FeNi is an attractive candidate as \emph{gap magnet}, i.e., a magnet with an expected energy product in between that of cheap ferrite magnets and that of the rather expensive high-performance magnets of the Nd-Fe-B family~\cite{COEY2012524}. 

The L$1_0$-ordered phase of FeNi was first reported by N\'eel and coworkers, who irradiated disordered Fe$_{50}$Ni$_{50}$ specimens with neutrons in the presence of magnetic field~\cite{Pauleve1962, Neel1964}. Later it was naturally observed in iron meteorite samples~\cite{PETERSEN1977192, ALBERTSEN1978, Danon1979, Danon1980, clarke}. The laboratory synthesis of the ordered phase is, however, very challenging because of the rather low order-disorder transition temperature, which is around 320\degree C~\cite{Pauleve1962, Neel1964, Reuter1989}. At such temperatures, atomic diffusion is extremely slow, which prevents the formation of the ordered state by conventional annealing techniques on realistic time-scales. Even though the synthesis of fully ordered L$1_0$ FeNi is very challenging, samples with a high degree of chemical order have been prepared in experiments using methods such as, e.g., nitrogen insertion and topotactic extraction \cite{Nitrogen-insertion}, the transformation from an amorphous state to a stable crystalline state \cite{Amorphous-to-crystalline}, pulsed laser deposition \cite{PLD}, and molecular beam epitaxy~\cite{Fe-Ni-composition}.

Due to the difficulty in obtaining fully ordered samples, it is important to understand the effect of partial chemical disorder on the magnetic properties of FeNi, in particular how deviations from equiatomic stoichiometry and perfect chemical order affect the magneto-crystalline anisotropy energy (MAE). Note that deviations from equiatomic stoichiometry will automatically result in a decrease of chemical order.
The compositional dependence of the MAE has been studied experimentally in partially ordered Fe-Ni thin films grown by molecular beam epitaxy, and a maximum of the MAE has been found for a composition of 60\,\% Fe~\cite{Fe-Ni-composition}.
However, an increase of the MAE with increasing Fe content relative to the fully ordered equiatomic case could not be confirmed by electronic structure calculations using the coherent potential approximation~\cite{Edstrom_et_al:2014}.

In previous work, we have used first principles calculations within density functional theory (DFT) to investigate the effect of partial chemical disorder on the MAE of equiatomic L$1_0$ FeNi~\cite{Izardar_2020}. By considering supercells with different distributions of Fe and Ni atoms over the available lattice sites, we have shown that the average MAE remains nearly constant if deviations from the perfect chemical order are not too large. Even a reduction in the degree of chemical order by about 25\,\% does not lead to a significant reduction of the MAE. This is very promising for the experimental synthesis of L$1_0$ FeNi with high MAE. 
Furthermore, several (randomly created) configurations with only partial chemical order exhibited a larger MAE than the perfectly ordered case. This raises the question of what factors determine the MAE in L$1_0$ FeNi, how it depends on the specific atomic distribution and composition, and whether the MAE can be further increased by optimizing the distribution of Fe and Ni atoms.  

In this article, we analyze the relation between the MAE and the local atomic environment, in order to identify ways to increase the MAE of FeNi beyond that of the perfectly L$1_0$-ordered equiatomic case. Thereby, we use the orbital magnetic moment anisotropy (OMA) as a proxy for a potential local contribution to the MAE. 
We first analyze 50 equiatomic configurations with 75\,\% chemical order, and identify a local nearest neighbor environment for the Fe atoms that appears to be particularly favorable for obtaining a high OMA. Using this favorable environment as a guide, we manually construct several new configurations with potentially high MAE and Fe contents between 50\% and 62.5\%. DFT calculations are then used to confirm the high MAE of these structures. 

We then perform a more systematic analysis using \emph{smooth overlap of atomic positions} (SOAP) as descriptors for the local atomic environment~\cite{SOAP}. This analysis indicates only a weak correlation between the local OMA and the local environment. Furthermore, this correlation exists only if the underlying structure does not deviate too much from perfect chemical order. This suggest that it is unlikely that the MAE in L$1_0$ FeNi can be understood fully as a sum of local contributions determined by the local environment.
We thus refrain from further analyzing the physical origin of the enhanced anisotropy in our specifically designed configurations. Instead, we explore the effect of randomly inserting excess Fe atoms into the Ni planes of the perfectly ordered structure (shown in Fig.~\ref{img:figures}b). This is motivated by the fact that the previously identified ``favorable environment'' implies a certain degree of Fe overstoichiometry. We find that, even though the excess Fe decreases the degree of chemical order, the MAE nevertheless increases, at least up to an Fe content of 62.5\,\%, consistent with previous experimental reports~\cite{Fe-Ni-composition}.
Our results thus identify a very promising route for optimizing the MAE using experimental layer-by-layer deposition techniques.

In the remainder of this article, we first describe our computational methods and how we model the partial chemical order (Sec.~\ref{sec:computational_details}), before we present the identification of the favorable environment (Sec.~\ref{sec:A}) and the specifically designed configurations with high MAE (Sec.~\ref{sec:B}). Sec.~\ref{sec:SOAP} contains the SOAP analysis and Sec.~\ref{sec:random structures from Pz equal to 1} the results for the structures with excess Fe atoms inserted into the Ni planes. Finally, our conclusions are summarized in Sec.~\ref{sec:Summary}.

\section{\label{sec:computational_details}Computational Method}

\subsection{Modeling of partially ordered structures}

\begin{figure}
  \centering
  \includegraphics[width=1.0\columnwidth]{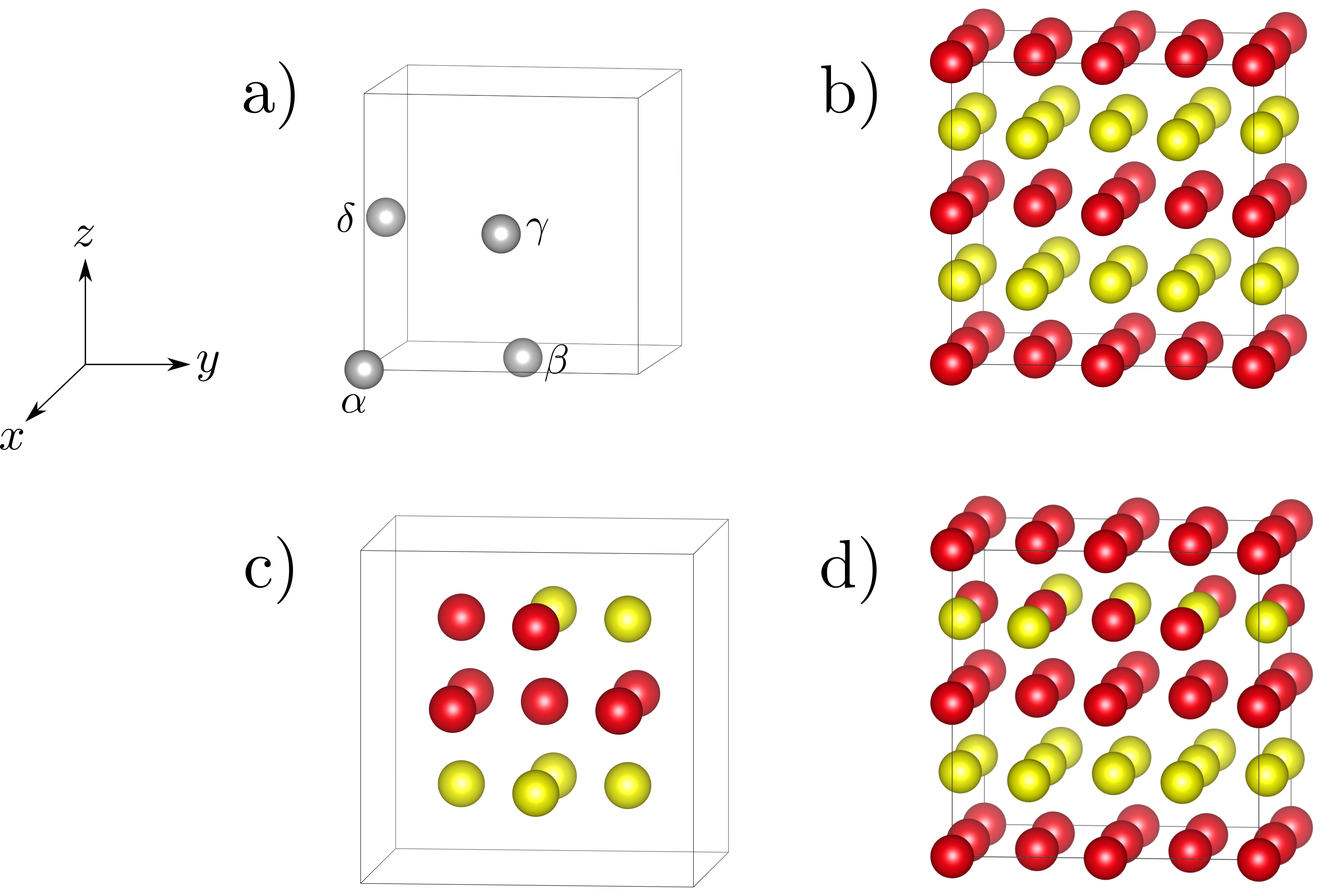}
  \caption{a) Depiction of the four sites of the fcc lattice within the conventional cubic unit cell, defining the four sublattices $\alpha$, $\beta$, $\gamma$, and $\delta$ used in \pref{eq:order-parameter}. b) Perfectly L$1_0$-ordered structure, depicted in a 2 $\times$ 2 $\times$ 2 supercell relative to the conventional four-atom cubic cell. c) Schematic representation of the ``favorable'' nearest neighbor environment of Fe atoms with the largest OMA of 0.015\,$\mu_\text{B}$. d) Specifically designed structure (with composition Fe$_{0.625}$Ni$_{0.375}$) based on this favorable environment. Red and yellow spheres in all subfigures represent Fe and Ni atoms, respectively.}
  \label{img:figures}
\end{figure}
 
To study different configurations with varying composition Fe$_x$Ni$_{1-x}$ with $0.5\leq x\leq 0.625$, we use a $2 \times 2 \times 2$ supercell of the conventional cubic cell (with lattice constant $a=3.56$\,\AA), containing 32 sites of the underlying fcc lattice (see also Ref.~\onlinecite{Izardar_2020}). We then occupy the lattice sites with Fe and Ni atoms in different ways to create structures with a certain Fe content and degree of order. To characterize the degree of order of a given configuration, we define the three-component long range order parameter $P_x$, $P_y$ and $P_z$ as follows:
\begin{equation}
\begin{aligned}
\label{eq:order-parameter}
P_x & = \tfrac{1}{2} \left( p_{\alpha} - p_{\beta} + p_{\gamma} - p_{\delta} \right) \\
P_y & = \tfrac{1}{2} \left( p_{\alpha} - p_{\beta} - p_{\gamma} + p_{\delta} \right) \\
P_z & = \tfrac{1}{2} \left( p_{\alpha} + p_{\beta} - p_{\gamma} - p_{\delta} \right)
\end{aligned}
\end{equation}
Here, $p_i$, $i \in \{ \alpha, \beta, \gamma, \delta \}$, denote the fraction of sites  
(or the occupation probabilities in the thermodynamic limit) for each sublattice $i$ that are occupied with Fe, and the different sublattices correspond to the four sites defining the fcc lattice within the conventional cubic cell (see Fig.~\ref{img:figures}a). The definition in \pref{eq:order-parameter} is consistent with the definition used in our previous work for equiatomic composition~\cite{Izardar_2020}, where $\sum_i p_i = 2$, but is also applicable to different Fe contents.
As can be seen in Fig.~\ref{img:figures}b, perfect L1$_0$ order consists of (001)-type planes that are alternatingly occupied with either Fe or Ni. Depending on whether these planes are stacked along the $x$, $y$, or $z$ direction, the corresponding component of the order parameter becomes non-zero.
In the following, we will, without loss of generality, always consider structures where the main component of the order parameter is oriented along the $z$ direction, i.e., $|P_z| \geq 0$, whereas $P_x$ and $P_y$ are either kept to zero or they average out over different configurations with the same $P_z$.

The MAE is defined as the energy difference for orientation of the magnetization along the order parameter direction ($z$ axis) and perpendicular to it (along either the $x$ or the $y$ direction). Note that for an arbitrary individual configuration, the $x$ and $y$ directions are in general not equivalent. In cases where we are interested in the MAE of an individual configuration, we therefore evaluate the MAE with respect to both $x$ and $y$ directions and then take the average, i.e., $\text{MAE}=(\text{MAE}_x+\text{MAE}_y)/2$ with MAE$_x=E^{[100]} - E^{[001]}$ and MAE$_y=E^{[010]} - E^{[001]}$. Here, $E^{[100]}$, $E^{[001]}$, and $E^{[010]}$ are the total energies obtained for magnetization aligned along the $[100]$, $[001]$ and $[010]$ directions, respectively. 
In cases where we average the MAE over different randomly created structures with the same $P_z$, we only calculate MAE$_x$, since the difference between the $x$ and $y$ directions will average out over many configurations.
In our definition, a positive MAE indicates that the easy axis is oriented along the order parameter direction, i.e., the $[001]$ direction, which is known to be the case for L$1_0$ FeNi. 

Note that for all configurations considered in this work, we keep the lattice constant and atomic positions fixed to those of an ideal fcc lattice with a $c/a$ ratio of 1.0. This allows to focus on the purely chemical effect related to the distribution of atoms over the available sites, and also significantly reduces the computational effort. Note that small changes in lattice parameters, e.g., resulting from a varying Fe content, are not expected to have a significant effect on the MAE. For example, we have verified that relaxing the $c/a$ ratio for the fully ordered configuration, changes the MAE by only $\sim 2 \mu$eV/(2 atoms).

\subsection{Computational details}

We calculate the  MAE from DFT calculations including spin-orbit coupling, using the Vienna \textit{ab} \textit{initio} Simulation package (VASP) \cite{Kresse1996}, the projector-augmented wave method (PAW) \cite{PAW1994,Kresse1999}, and the generalized gradient approximation according to Perdew, Burke, and Ernzerhof~\cite{PBE}. To obtain the MAE, we use the magnetic force theorem~\cite{Weinert/Watson/Davenport:1985,Daalderop/Kelly/Schuurmans:1990},
i.e., we perform non-self-consistent calculations with spin-orbit coupling included, using the charge density converged without spin-orbit coupling, and then take the difference in band energies between the two different orientations of the magnetization direction. Our PAW potentials include 3$\textit{p}$, 4$\textit{s}$, and 3$\textit{d}$ states in the valence for both Fe and Ni. Brillouin zone integrations are performed using the tetrahedron method with Bl\"{o}chl corrections and a $\Gamma$-centered $14\times14\times14$ \textbf{k}-point mesh. The plane wave energy cut-off is set to 350\,eV, and the total energy is converged to an accuracy of $10^{-8}$\,eV. The convergence of the MAE with respect to the \textbf{k}-point sampling was tested by performing calculations using up to $25\times25\times25$ \textbf{k}-points. Thereby, the MAE was found to be sufficiently converged for our purposes, to about $\pm 1\,\mu$eV per 2 atoms, using a $14\times14\times14$ \textbf{k}-point mesh. 

To parameterize the atomic distribution within the local environment around each Fe atom, we use the SOAP descriptor~\cite{SOAP} implemented in the DScribe software package \cite{HIMANEN2020106949}. 
In this approach, the distribution of atoms around a specific site is represented by a superposition of Gaussians. The resulting atomic density is then expressed in terms of spherical harmonics and suitable radial basis functions. From the resulting expansion coefficients, a rotationally invariant \emph{power spectrum} is obtained~\cite{SOAP}, which we refer to in the following as ``SOAP vector''.
We employ cutoff radii for the local environment of 2.6\,\AA, 3.6\,\AA, and 4.4\,\AA, corresponding to first nearest neighbor (1NN), up to second nearest neighbor (2NN), and up to third nearest neigbor (3NN) environments, respectively. We use a basis of $n_\text{max} = 8$ radial polynomial functions (as defined within DScribe) and $l_\text{max} = 6$ angular functions, and a Gaussian width of 0.5\,\AA. 

We then use t-distributed stochastic neighbor embedding (t-SNE)~\cite{JMLR:v9:vandermaaten08a}, as implemented in scikit-learn \cite{scikit-learn}, to reduce the high-dimensional SOAP vectors to two dimensions for easier visualization and analysis. 
This technique generates a low-dimensional map of the original data points by constructing pairwise probability distributions in both the high- and low-dimensional space, such that similar data points are assigned a high joint probability, and then finds the positions in the low-dimensional space by minimizing the difference between the two probability distributions.
As a result, similar data points are mapped onto points that are close to each other, potentially forming clusters in the low dimensional representation, while dissimilar data points are preferentially mapped onto more distant points.
Thereby, the size of the clusters in the low-dimensional map is strongly influenced by the ``perplexity'', the main parameter determining the mapping, which can roughly be viewed as a measure for the effective number of neighboring points in the high dimension. We set the perplexity to 6 (and use default values for all other parameters). Larger values lead to larger clusters and thus less separation between data points with different OMA, which does not affect our conclusions.
Note that we simply use the Euclidean distance between the SOAP vectors to define similarity in the high dimensional space.


\section{\label{sec:results}Results and Discussion}
\subsection{\label{sec:A}Local atomic environment of Fe atoms}

In our previous work~\cite{Izardar_2020}, we calculated the MAE for 50 configurations with $P_z=0.75$ for the equiatomic composition (Fe$_{0.5}$Ni$_{0.5}$). Many of these configurations exhibited a MAE larger than that of the fully ordered structure (76 \,$\mu$eV/2 atoms), with one configuration even exceeding an MAE of 125\,$\mu$eV/2 atoms. This raises the question of what factors determine the MAE in FeNi and how exactly it depends on the specific distribution of Fe and Ni atoms. 
To address this question, we now assume that the MAE can be understood, at least approximately, as a sum of local contributions that are determined by the local environment of the individual atoms in the system. We then use the OMA, i.e., the difference in the orbital magnetic moment for different orientations of the magnetization relative to the crystal axes, as a proxy for such a local contribution to the MAE.
As shown by Bruno and others~\cite{Bruno_1989,Laan:1998}, using a perturbative treatment of the spin-orbit coupling within the tight binding approximation, the MAE is proportional to the OMA under certain conditions. Indeed, previous work on L$1_0$ FeNi has found a correlation between MAE and the (site-averaged) OMA both for the fully ordered state~\cite{Miura_2013} as well as for varying degrees of order~\cite{Izardar_2020} and also showed that the OMA is dominated by the contribution of the Fe atoms.

\begin{figure}
   \centering
   \includegraphics[width=0.9\columnwidth]{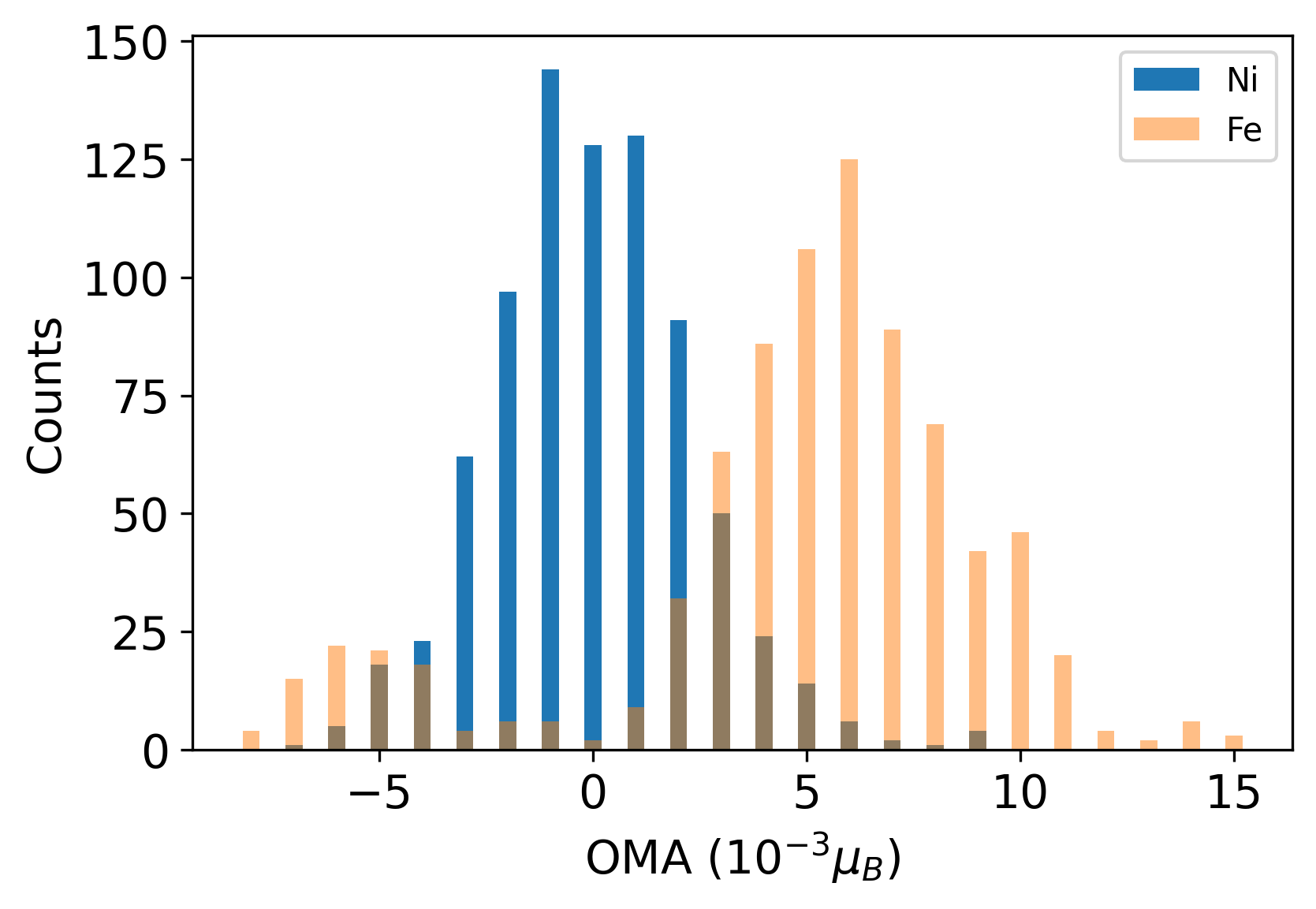}
   \caption{Distribution of the individual OMA values (in $10^{-3}\mu_\text{B}$), obtained for all Fe and Ni atoms included in the 50 equiatomic configurations of FeNi with long range order parameter $P_z=0.75$.}
   \label{img:OMA_hist_75}
\end{figure}

The latter can also be seen from Fig~\ref{img:OMA_hist_75}, which shows the distribution of the OMA for the individual Fe and Ni atoms in all 50 configurations with $P_z=0.75$. Here, we define the OMA as $\Delta L=L^{[001]} - L^{[100]}$, where $L^{[001]}$ and $L^{[100]}$ are the orbital magnetic moments obtained when the total magnetization is oriented along $[001]$ and $[100]$ directions, respectively. We do not consider the $[010]$ direction, since all 50 configurations are designed to have $P_x=P_y=0$ and thus the anisotropy between the $x$ and $y$ direction is assumed to be small. Note that a positive OMA indicates that the orbital magnetic moment is larger if the magnetization is oriented along the easy axis, i.e., along the $[001]$ direction.
The distribution of the OMA for the Fe atoms in Fig.~\ref{img:OMA_hist_75} is asymmetric around zero, with the majority of Fe atoms exhibiting a positive OMA. In contrast, the distribution for the Ni atoms is rather symmetric, with a peak centered around zero OMA. This confirms that the total OMA is dominated by the Fe contribution, consistent with previous first principles calculations~\cite{Izardar_2020, Miura_2013}, and also magnetic circular dichroism measurements which found a strong angular dependence of the orbital moments for Fe but not for Ni~\cite{kotsugi2013}. 
In the following, we therefore focus on analyzing only the local environment of the Fe atoms.

From Fig.~\ref{img:OMA_hist_75}, one can also see that a small number of Fe atoms exhibit a particular high OMA of 0.015\,$\mu_\text{B}$. Analysis of their local atomic environment reveals that all these Fe atoms exhibit an equivalent first nearest neighbor (1NN) environment, which is shown in Fig.~\ref{img:figures}c. Each central Fe atom is surrounded by six Fe and six Ni atoms, with four Fe atoms in the same $x$-$y$ plane and the remaining two Fe atoms as nearest neighbors to each other in the plane either above or below the central Fe atom. All other 1NN positions are occupied by Ni atoms.
There are also some other Fe atoms in our 50 configurations that have exactly the same (or an equivalent) 1NN environment as depicted in Fig.~\ref{img:figures}c but exhibit an OMA smaller than 0.015 $\mu_\text{B}$. 
This indicates that the OMA (and thus likely the MAE) is not completely determined by only the 1NN environment, but that further neighbors also play a role (a further analysis of this is presented in \pref{sec:SOAP}). 
Nevertheless, most Fe atoms with this specific 1NN environment exhibit a OMA higher than that of the fully ordered case (i.e., higher than 0.006 $\mu_\text{B}$), and from now on we will therefore refer to this as the ``favorable environment''.

\subsection{\label{sec:B}New configurations based on the favorable local environment}

Next, we use the favorable environment as a guide to design optimized configurations with potentially high OMA, and thus high MAE. We start with a $2 \times 2 \times 2$ supercell of the conventional cubic cell and place Fe atoms at positions $(0, 0, 0)$ and $(0.5, 0.5, 0.5)$. We then arrange Fe and Ni atoms on the corresponding 1NN positions of the two initial Fe atoms according to the favorable environment. The remaining six positions in the supercell are then filled with Fe atoms, thereby maximizing the number of Fe atoms exhibiting the favorable environment. This procedure results in a configuration with 20 Fe and 12 Ni atoms ($x=0.625$), depicted in Fig.~\ref{img:figures}d,  where 16 Fe atoms exhibit the favorable environment. The corresponding long-range order parameter is $P_x=P_y=0$ and $P_z=0.75$.
Note that the Fe:Ni ratio on the 12 1NN-sites of the favorable environment is 6:6, whereas it is 4:8 for the fully ordered equiatomic case. This results in an excess Fe content of the specifically designed structure, with 20 Fe and 12 Ni atoms in the unit cell.

We now use DFT to calculate the MAE of this specifically designed structure and obtain a value of 141\,$\mu$eV/2 atoms. This value is nearly a factor of two higher than the MAE of 76\,$\mu$eV/2 atoms obtained for the perfectly ordered structure. Further analysis shows that the 16 Fe atoms with the favorable environment exhibit an OMA of 0.008 $\mu_B$, consistent with the analysis in \pref{sec:A}. The remaining four Fe atoms have an OMA of only 0.001\,$\mu_B$ while the 12 Ni atoms exhibit small OMA ranging between $-0.002$ and 0.003\,$\mu_B$. The total OMA of the 32-atom supercell is 0.144\,$\mu_B$, which is also much higher than that of the perfectly ordered structure (0.064\,$\mu_B$). It thus appears that optimizing the local 1NN environment of the Fe atoms can indeed lead to configurations with particularly high MAE.

\begin{figure}
   \centering
   \includegraphics[width=\columnwidth]{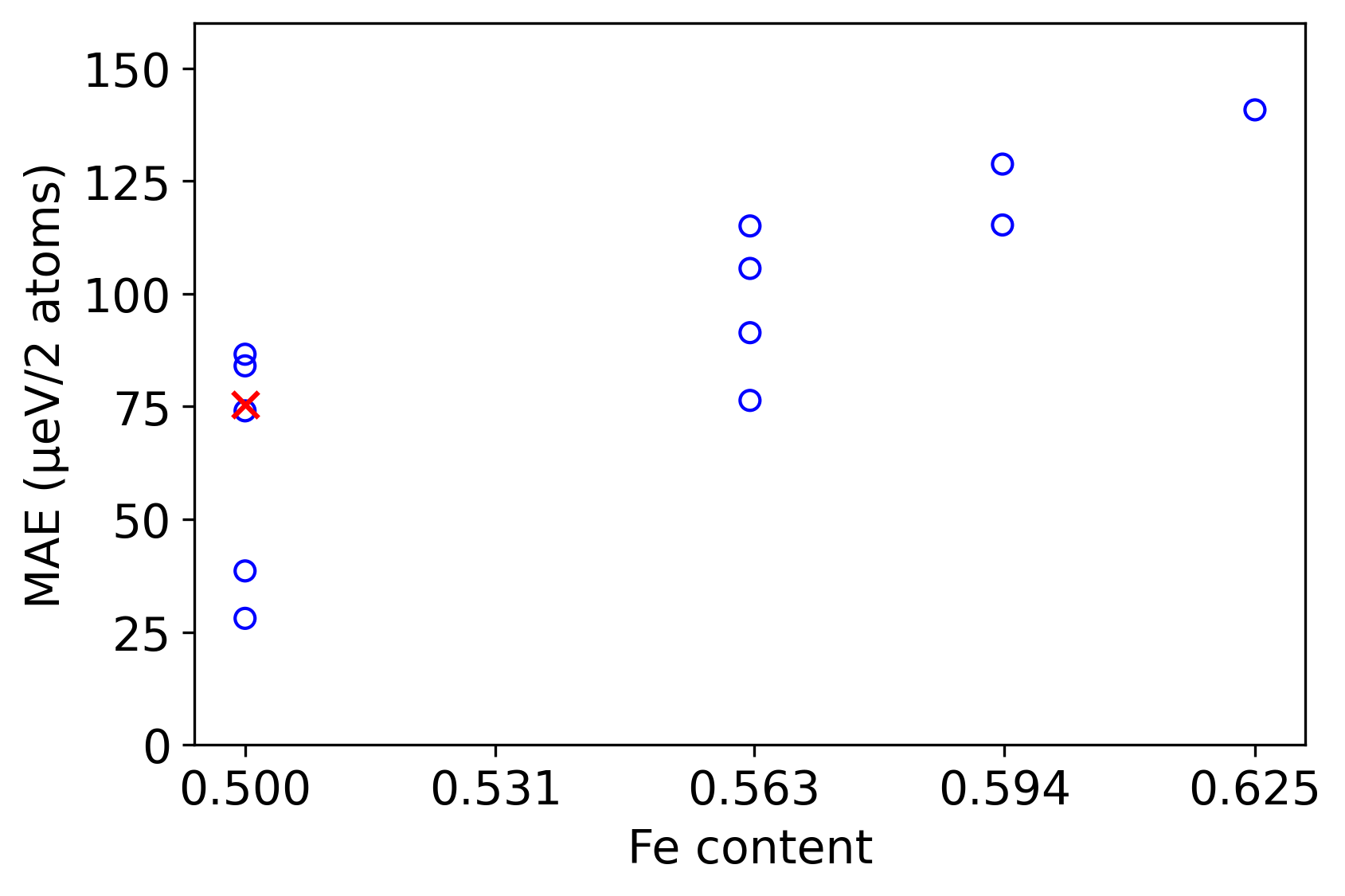}
   \caption{MAE calculated for different configurations with varying Fe content, obtained by randomly replacing some Fe atoms with Ni starting from the optimized structure with 20 Fe atoms ($x=0.625$). The red cross represents the MAE of the fully ordered structure.} 
   \label{fig:MAE_hands}
\end{figure}

Motivated by this encouraging result, we create additional structures with Fe to Ni ratios of 19:13 ($x=0.594$), 18:14 ($x=0.563$), and 16:16 ($x=0.5$), by starting from the optimized 20:12 structure and then successively replacing one or more Fe atoms by Ni (without further optimization of the local environment). The calculated MAE for these structures as well as or the fully ordered structure are shown in \pref{fig:MAE_hands}. 
It can be seen that, compared to the optimized structure with 62.5\,\% Fe, the MAE decreases with decreasing Fe content while simultaneously the number of Fe atoms exhibiting the favorable environment is reduced to 11 and 8 for $x=0.594$, and to 6, 8, 6, and 0 for $x=0.563$ (listed from highest to lowest MAE in each case). However, all structures with Fe content larger than 0.5 exhibit a MAE that is higher than that of the perfectly ordered equiatomic structure. For the equiatomic composition, some structures also exhibit a relatively low MAE, and the number of Fe atoms exhibiting the favorable environment in this case varies between 0 and 4. Thus, the corresponding structures cannot be viewed any more as specifically optimized with respect to the local environment.

Nevertheless, this simple analysis indicates that it is in principle possible to enhance the MAE in L$1_0$ FeNi by moving to slightly Fe rich compositions and optimizing the distribution of Fe and Ni atoms, and confirms our previous result that the perfectly ordered equiatomic state is not the one with the highest MAE. However, in order to obtain guidelines on how to potentially achieve the corresponding structures in an experimental synthesis, a more detailed understanding of the relationship between atomic configuration and the resulting MAE is required.

\subsection{\label{sec:SOAP}Correlation between OMA and the local chemical environment}

\begin{figure*}
   \centering
   \includegraphics[width=0.8\textwidth]{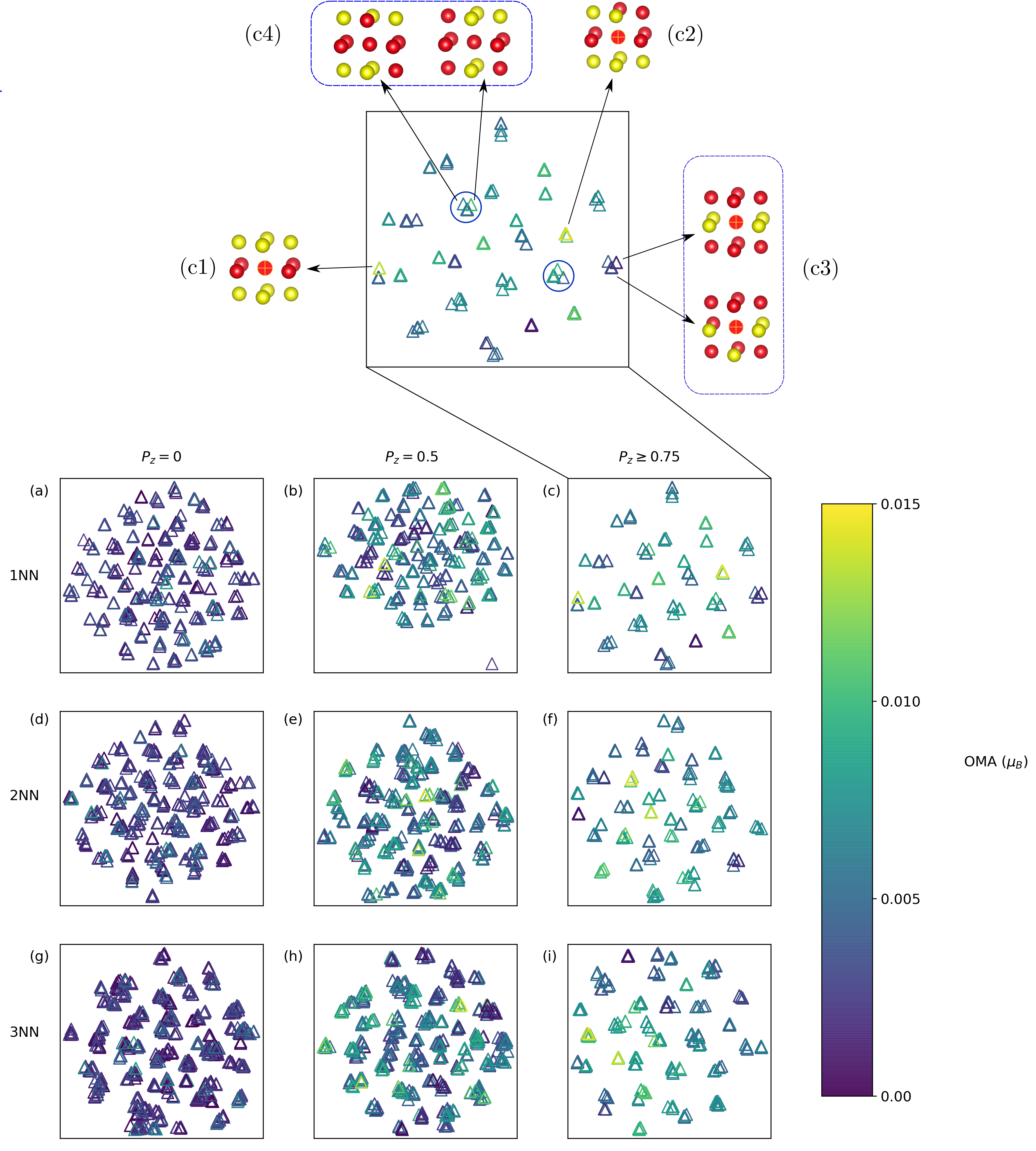}
   \caption{The t-SNE visualization of the SOAP vectors representing different local chemical environments of Fe atoms including up to 1NN, [(a)-(c)], 2NN [(d)-(f)], and 3NN [(g)-(i)] environments corresponding to configurations with $P_z = 0.0$, $P_z = 0.5$, and $P_z \geq 0.75$. The axes of the subplots represent the reduced dimensions of the SOAP vectors.  The color indicates the corresponding absolute value of the OMA. The zoomed version of subplot (c) depicts different local atomic environments (1NN only) corresponding to selected data points.}
   \label{img:SOAP}
\end{figure*}

To obtain a deeper insight into the correlation between the local atomic environment of an Fe atom and its OMA, it is desirable to have a more systematic and quantitative way to characterize the distribution of atoms within the different local atomic environments. For this purpose, we employ the smooth overlap of atomic positions (SOAP) approach~\cite{SOAP}, which encodes the local atomic structure around an atom up to a specified cutoff distance in the form of well-defined \emph{descriptors}.
The ``SOAP vectors'' provide a rotationally invariant representation of the different local chemical environments, i.e., an abstract parameterization of  the spatial distribution of all atoms up to the cutoff distance, around each site.  
We generate SOAP vectors corresponding to local atomic environments including up to first-nearest neighbors (1NN), second-nearest neighbors (2NN), and third-nearest neighbors (3NN) for all Fe atoms in the various configurations.

For better analysis and visualization, the high-dimensional SOAP vectors need to be projected into two dimensions in a way that preserves the relative distances between the SOAP vectors as much as possible. Note that the ``distances'' between different SOAP vectors are a measure of how similar or dissimilar the corresponding chemical environments are (albeit in a not necessarily intuitive way).
As briefly outlined in Sec.~\ref{sec:computational_details}, we use t-SNE to map each high-dimensional SOAP vector into a two dimensional data-point in such a way that similar vectors, i.e. SOAP vectors representing similar local chemical environments, are modeled by data-points that are near to each other, while dissimilar vectors, i.e., vectors representing different local chemical environments, are modeled by points that are far away from each other.
Thus, similar atomic environments will form clusters in the low-dimensional representation, while atomic environments that are very different from each other will be separated into different clusters.
However, we note that the specific distances between different clusters in the low-dimensional representation are not meaningful, as t-SNE mainly preserves the \emph{local} similarity structure of the data while mapping objects from high to low dimensions. 

Fig.~\ref{img:SOAP} shows the t-SNE visualization of the SOAP vectors representing different local atomic environments, including up to 1NN, 2NN, and 3NN, for Fe atoms in different equiatomic configurations with different $P_z$. For each value of the long range order parameter, 50 configurations have been randomly generated.
The $x$ and $y$ axes of the subplots represent the reduced dimensions of the SOAP vectors obtained using the t-SNE technique, while the symbols are colored according to the \emph{absolute value} of the OMA of the corresponding Fe atoms. 
We note that a potential correlation between the local environment and the local OMA implies that, if two local environments are mapped onto each other by some rotation, then the corresponding anisotropies should be rotated accordingly. We therefore relate only the absolute values of the OMA to the corresponding SOAP vectors. Nevertheless, since we are not evaluating the OMA relative to the $y$-direction, a data-point with very low OMA in Fig.~\ref{img:SOAP} could still exhibit a strong uniaxial anisotropy along $y$, which represents a certain limitation of our analysis. However, at least for $P_z=0.75$, and since $P_x=P_y=0$ was imposed for all configurations, such cases can be considered as unlikely.

We first consider the case with $P_z \geq 0.75$ with only the 1NN included [Fig.~\ref{img:SOAP}(c)].
The zoomed version of  Fig.~\ref{img:SOAP}(c) shows some examples of the local 1NN environment of the Fe atoms corresponding to selected data points. Case (c1) corresponds to the local atomic environment found in the perfectly ordered structure, while the cluster of yellow and light green triangles denoted as (c2) corresponds to the previously identified ``favorable" environment with the highest OMA (see Sec.~\ref{sec:A}). Note that due to the small truncation error in the SOAP expansion and other numerical inaccuracies, the corresponding data-points (yellow and light green colored triangles) are slightly separated in the t-SNE representation, even though they in fact correspond to identical atomic environments. However, we have carefully verified that such small numerical differences in the SOAP vectors do not lead to an artificial separation of equivalent points into different clusters. 

For another cluster, (c3), containing only dark blue triangles indicative of a very small OMA, one can see that one of these configurations corresponds to an atomic environment that resembles an Fe atom within a nominal Ni plane, and that the other configuration that is shown indeed exhibits a nearly identical environment, but with one Fe-Ni pair exchanged between the lower Fe plane and the nominal Ni plane in the middle.

One can also identify some clusters in Fig.~\ref{img:SOAP}(c) that contain a range of colors, indicating Fe atoms with similar 1NN environment but rather different values of the OMA [e.g., the two clusters that are encircled in the zoomed version]. Case (c4) explicitly shows the chemical environment for two such cases. Indeed both cases represent a similar chemical environment corresponding to an Fe atom within a nominal Fe plane with two or three additional Fe atoms distributed in the planes above and below. 

On including further neighbors into the local atomic environment, up to 2NN (Fig.~\ref{img:SOAP}(f)) and 3NN (Fig.~\ref{img:SOAP}(i)), the coloring of data points within individual clusters (and thus the correlation between OMA and atomic environment) seems to become more consistent~\footnote{Note that including further neighbor shells fundamentally changes the nature of the chemical environment, and thus the corresponding SOAP vectors as well as their positions in the two-dimensional map change inevitably.}.
However, even when considering up to 3NN, the correlation between chemical environment and OMA is still not perfect.
This indicates that one would need to go even beyond 3NN to establish a clear correlation between a specific atomic environment and the corresponding OMA of the central Fe atom. 

For configurations with $P_z = 0.5$, one can see that the correlation between the local atomic environment of the Fe atoms and their OMA becomes even weaker, even when including up to 3NN [indicated by several clusters containing both light green and dark blue triangles in Fig.~\ref{img:SOAP}(h)]. 

For $P_z = 0.0$ [Fig.~\ref{img:SOAP}(a), (d), and (g)], the picture is less clear, since there are only few data points with a large OMA. However, it seems that many clusters contain the full spectrum of available OMA values (from dark blue to light green), with no systematic improvement on including further neighbors in the atomic environment.

We conclude that in general (i.e., independent of the global configuration and the resulting value of the long range order parameter $P_z$) the OMA is not uniquely determined by the local environment, even if including up to 3NN. It is therefore unlikely that the MAE of partially ordered FeNi can be efficiently described by a simple model based on a sum of local contributions determined mainly by the local environment of the Fe atoms. 
This seems also consistent with Ref.~\onlinecite{Ke:2019}, where the MAE of fully ordered L$1_0$ FeNi was decomposed in inter- and intra-atomic contributions using Wannier functions, and it was found that the inter-atomic term is dominant, questioning the applicability of a single ion model of the MAE for FeNi. 

Thus, even though the manual identification of the ``favorable environment'' has allowed to successfully design configurations with high MAE in Sec.~\ref{sec:B}, the analysis in Fig.~\ref{img:SOAP} shows that there is only a partial correlation between OMA and the 1NN environment. Furthermore, even if considering up to 3NN, this partial correlation seems to only hold for $P_z \geq 0.75$, i.e., when a high degree of chemical order (and our limited supercell size) introduces an implicit constraint on the available further neighbor configurations.

\subsection{\label{sec:random structures from Pz equal to 1} Perfectly ordered structure doped with excess Fe}

The SOAP analysis in Sec.~\ref{sec:SOAP} shows that it is unlikely that the MAE in FeNi can be understood within a simple model based on the local atomic environment of individual atoms. Furthermore, even if such a simple model could explain the high MAE of our specifically designed structure based on the favorable environment in Sec.~\ref{sec:B}, it would be virtually impossible to synthesize the corresponding structure experimentally. Here, we therefore pursue a different route to further optimize the MAE in FeNi. It is based on the observation that the favorable environment implies an excess Fe content compared to the perfectly ordered equiatomic case, and that the specifically designed structure with high MAE shown in Fig.~\ref{img:figures}(d) essentially corresponds to the perfectly ordered case, but with some additional Fe atoms located in the nominal Ni planes. In the following, we therefore investigate how a random introduction of excess Fe atoms into the nominal Ni planes affect the MAE.
Such configurations could be obtained experimentally either by using layer-by-layer growth methods, or by maximizing the chemical order parameter in a structure with slightly more than 50\,\% Fe content.

We create configurations corresponding to different stoichiometries by randomly replacing Ni atoms with Fe starting from the perfectly ordered structure, i.e., we keep $p_\alpha=p_\beta=1$ in \pref{eq:order-parameter}, whereas $p_\gamma$ and $p_\delta$ become nonzero, such that $\sum_i p_i =4x$, according to the given stoichiometry. Note that this automatically leads to the highest possible value of the long range order parameter, $P_z=2(1-x)$, achievable for a specific Fe content with $x>0.5$. We consider up to 62.5\,\% Fe, which corresponds to 20 Fe atoms within our 32 atom supercell.

\begin{figure}[t]
   \centering
   \includegraphics[width=\columnwidth]{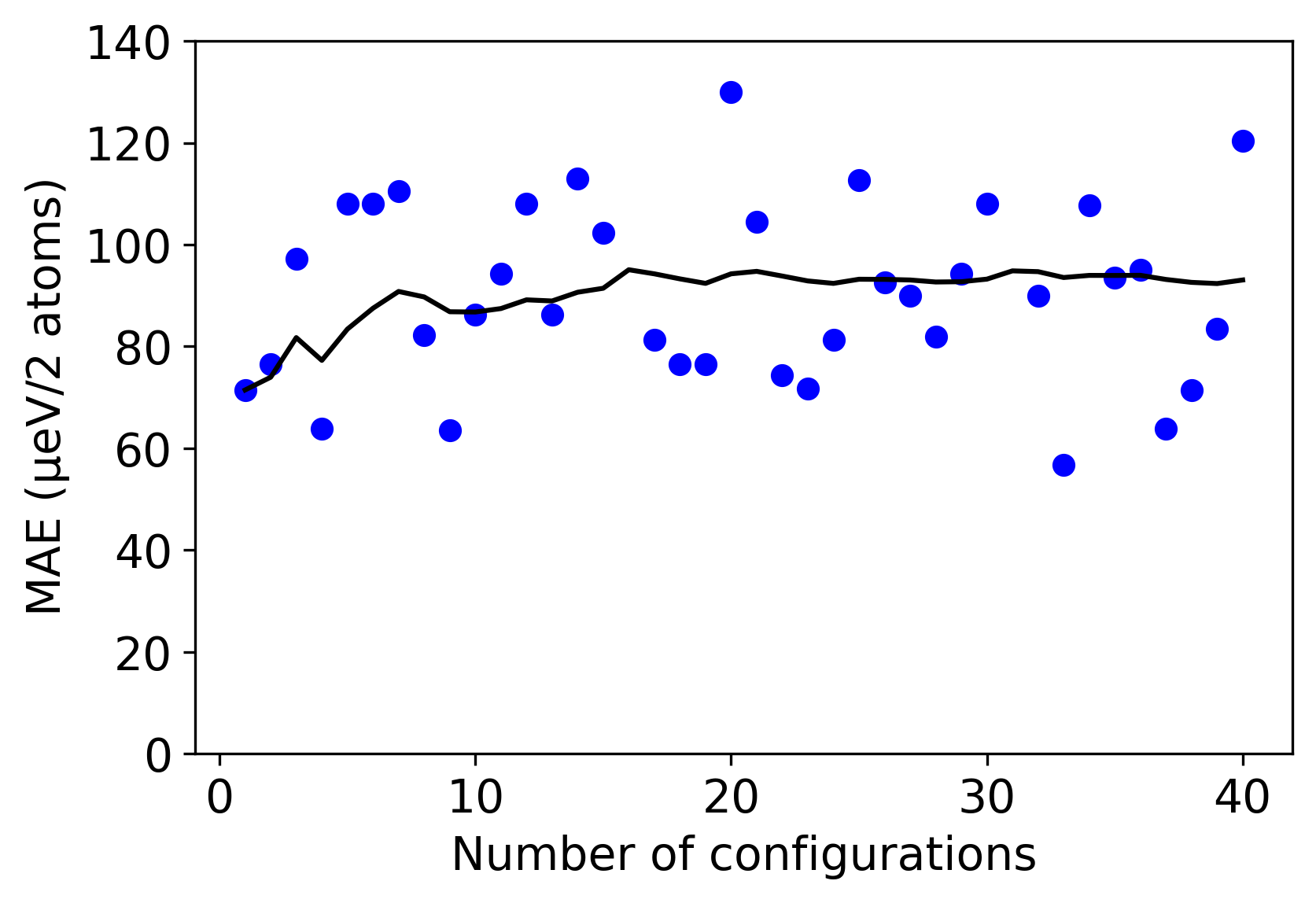}
   \caption{Calculated MAE values for 40 random configurations with composition belonging to Fe$_{0.625}$Ni$_{0.375}$ and a maximum long-range order parameter of $P_z=0.75$ (blue dots). The cumulative average is shown by the solid black line.}
   \label{Cumulative average}
\end{figure}

For compositions $x=0.531$, $x=0.563$, and $x=0.594$, corresponding to 17, 18, and 19 Fe atoms within the supercell, we create all possible configurations and then use XtalComp~\cite{XtalComp} to identify symmetry-equivalent structures. For these three specific stoichiometries, we find 1, 7, and 12 distinct groups of equivalent structures within the 32-atom supercell. We then calculate the MAE as average over MAE$_x$ and MAE$_y$ of one representative structure for each distinct group, and evaluate the average over all groups by considering the correct multiplicities.
For $x=0.625$, i.e., 20 Fe atoms in the supercell, the number of possible configuration becomes very high and we therefore do not perform a full symmetry analysis of equivalent structures.  
Instead, we sample the MAE for 40 randomly created configurations. Fig.~\ref{Cumulative average} shows the evolution of the cumulative average of the MAE for an increasing number of sampled configurations. One can see that the cumulative average converges to a constant value after about 30-40 configurations, which shows that we average over a sufficient number of configurations.

\begin{figure}[t]
   \centering
   \includegraphics[width=\columnwidth]{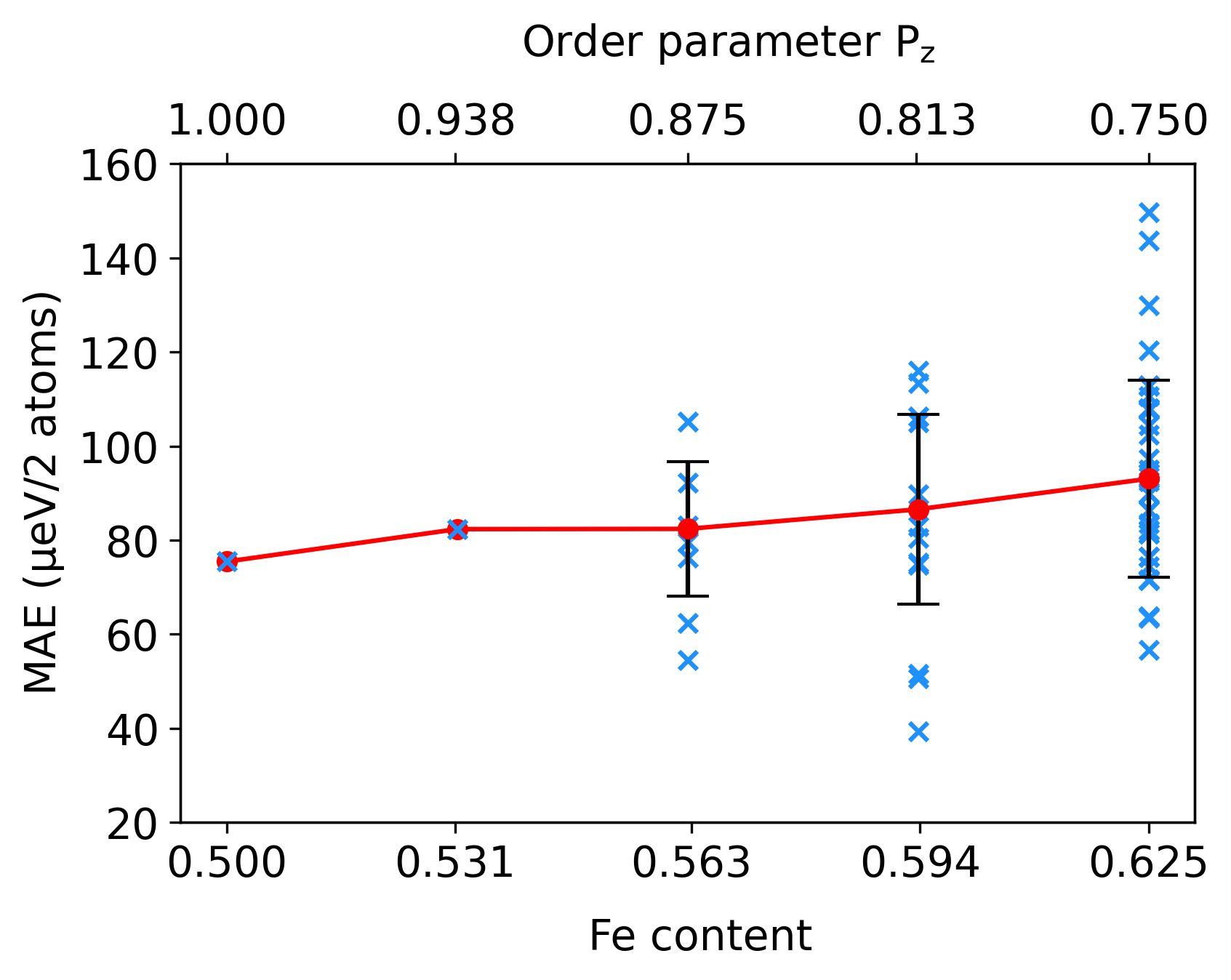}
   \caption{Calculated MAE as function of Fe content obtained by randomly replacing Ni atoms with Fe, starting from the perfectly ordered case. The corresponding reduction of the long range order parameter $P_z$ is indicated on the upper horizontal axis. The MAE values of the individual configurations for each Fe content are indicated by blue crosses, with the corresponding standard deviation marked by the black vertical bars. Corresponding average values are shown as connected red dots.}
   \label{MAE,Pz,and Fe content}
\end{figure}

Fig.~\ref{MAE,Pz,and Fe content} shows the resulting MAE as a function of Fe content. The corresponding reduction of the long range order parameter $P_z$ is also indicated.
Remarkably, the average MAE continuously increases with increasing Fe content, at least up to the highest considered Fe content of 62.5\,\%. For this case the average MAE is 93\,$\mu$eV/2 atoms, which corresponds to a $23$\,\% increase compared to the fully ordered equiatomic case.
Thus, it appears that the MAE of L$1_0$ FeNi can be increased by considering Fe-rich stoichiometries as long as the degree of chemical order can be kept high. This is consistent with the experimental results obtained by Kotsugi {\it et al.}~\cite{Fe-Ni-composition}, even though in their case the degree of chemical order was significantly lower than in our calculations.
We note that, for increasing Fe content, it will likely become significantly more difficult to maintain a high degree of chemical order in the system, and thus we do not consider Fe concentrations higher than 62.5\,\%.
Furthermore, the ferromagnetic ground state becomes more and more unstable towards the Fe-rich side, in favor of noncollinear antiferromagnetic or spin-spiral-type ordering (see, e.g., \cite{Abrikosov_et_al:2007,Lavrentiev_et_al:2014}).

\section{\label{sec:Summary}Summary and Conclusions}

In summary, we have investigated the effect of chemical disorder on the MAE in L$1_0$-ordered FeNi using first principles DFT calculations. Our calculations clearly show that the perfectly ordered equiatomic configuration is not the one with the highest MAE in this system. By analyzing the local OMA in a number of configurations with reduced chemical order, we were able to identify an apparently favorable 1NN environment, which then allowed us to design an optimized configuration with 62.5\,\% Fe content and a MAE nearly twice that of the fully ordered case.

However, further analysis using SOAP as descriptor for the local atomic environment, indicates that a purely local model might not be applicable for the MAE in FeNi, and thus developing a detailed understanding of how the MAE depends on the specific distribution of Fe and Ni atoms is rather challenging.
We note that it might still be possible, though, that the use of a different descriptor, or a different similarity measure for the chemical environment, could reveal a clearer correlation betwen OMA and local environment, compared to the SOAP analysis presented here.

In any case, our identification of the favorable environment implies that an increase of the MAE in partially ordered FeNi might be possible by inserting additional Fe atoms into the nominal Ni planes of the L$1_0$ structure.
Indeed, our corresponding DFT calculations confirm an increase of the MAE by nearly 25\,\% compared to the fully ordered case for an Fe content of 0.625. 
We note that our analysis presented in Fig.~\ref{img:OMA_hist_75} and also in our previous work, Ref.~\onlinecite{Izardar_2020}, indicates that the dominant contribution to the total OMA, and probably also to the MAE, stems from the Fe atoms. It therefore appears that incorporating more Fe atoms into the system can potentially increase the MAE as long as the degree of chemical order can be kept sufficiently high.
Thus, our results suggest a realistic route for optimizing the MAE in L$1_0$ FeNi using, e.g.,  layer-by-layer growth methods that allow to incorporate excess Fe atoms while keeping the degree of order as high as possible.

\begin{acknowledgments}
This work was supported by ETH Z\"urich. Calculations were performed on the cluster \enquote{Piz Daint}, hosted by the Swiss National Supercomputing Centre, and the \enquote{Euler} cluster of ETH Z\"urich.
\end{acknowledgments}

\bibliography{main}
\end{document}